# Elastic metamaterials with simultaneously negative effective shear modulus and mass density

Ying Wu[1,2], Yun Lai[1,3] and Zhao-Qing Zhang[1*]

[1] *Department of Physics, Hong Kong University of Science and Technology, Clear Water Bay, Kowloon, Hong Kong, China*

[2] *Division of Mathematical and Computer Sciences and Engineering, King Abdullah University of Science and Technology, Thuwal 23955-6900, Kingdom of Saudi Arabia*

[3] *Department of Physics, Soochow University, 1 Shizi Street, Suzhou 215006, People's Republic of China*

Email: phzzhang@ust.hk

## Abstract

We propose a type of elastic metamaterial comprising fluid-solid composite inclusions which can possess negative shear modulus and negative mass density over a large frequency region. Such a solid metamaterial has a unique elastic property that only transverse waves can propagate with a negative dispersion while longitudinal waves are forbidden. This leads to many interesting phenomena such as negative refraction, which is demonstrated by using a wedge sample, and a significant amount of mode conversion from transverse waves to longitudinal waves that cannot occur on the interface of two natural solids.

Recently, the artificial electromagnetic materials denoted metamaterials have attracted a great deal of attention due to their unusual properties. One example is the left-handed material (LHM), which was first predicted by Veselago [1]. LHM is a material with simultaneously negative permittivity and permeability, which leads to electromagnetic (EM) waves propagating with the wave vector opposite to the Poynting vector. LHM gives rise to many intriguing phenomena, such as negative refraction, reversed Doppler shift [1], and a perfect lens [2] etc. The first realization of LHM for EM waves consisting of metallic wires and split rings was proposed by Pendry *et al* [3] and fabricated by Smith *et al* [4]. Thereafter, an enormous progress has been made in realizing negative refraction with metamaterials at various frequency regimes [5-8].

Very recently, the principle of metamaterial has been extended to acoustic and elastic media, and various schemes have been proposed to realize acoustic and elastic LHM [9-13]. Different from EM and acoustic media, a normal elastic medium can propagate both longitudinal and transverse waves. In an isotropic solid, these waves are described by three independent material parameters, i.e., mass density, $\rho$, bulk modulus, $\kappa$, and shear modulus, $\mu$. Analogous to EM metamaertials, negative effective elastic parameters can be realized by introducing resonant structures into the building blocks of an elastic metamaterial. For example, due to dipolar resonances in locally resonant rubber-coated-lead structure [10, 14] and thin membranes [12, 15, 16], these structures have been shown to achieve negative mass density. Air bubbles in water [10] and Helmholtz resonator [11, 17] can achieve negative bulk modulus as results of monopolar resonances. It has also been found that negative shear modulus is related to quadrupolar resonance [18, 19]. But to the best of our knowledge, there has been no report of resonant structures to realize negative shear modulus with a noticeable bandwidth yet.

Finding negative effective shear modulus is not only of scientific interest, but also has intriguing applications. For instance, if an elastic LHM has simultaneously negative density and shear modulus in a certain frequency regime, it may give rise to negative refraction of transverse waves, i.e., $n_t = \sqrt{\rho}\sqrt{1/\mu} < 0$. Meanwhile, if the material's bulk modulus is positive so that $\kappa + \mu > 0$, the longitudinal wave becomes evanescent and the elastic LHM can be used as a polarizer. It will be shown later that total mode conversion from shear waves to longitudinal waves (and vice versa) can also be achieved when such a metamaterial is put in contact with a normal solid. This is not possible for any two natural solids. Here, we propose a new type of elastic metamaterial with a fluid-solid composite. Such a composite can support both dipolar and quadrupolar resonances in its building blocks and produce simultaneously negative effective mass density and negative effective shear modulus in a large frequency region, resulting in a wide negative band for shear waves and a gap for longitudinal waves. The effective medium parameters for $\rho$, $\kappa$ and $\mu$ can be obtained from an effective medium theory (EMT) [18]. The existence of a large negative band is confirmed by the band structure calculations using the multiple-scattering theory (MST) [20]. The negative refraction is demonstrated by MST simulations with a wedge sample. Interestingly, we find a significant amount of mode conversion from shear waves to longitudinal waves induced by the double-negative property of the metamaterial. Such a large conversion only occurs under negative refraction and does not occur in any natural solids.

The two-dimensional elastic metamaterial proposed in this work is composed of cylindrical fluid-solid composite inclusions arranged in a triangular lattice in the *x-y* plane embedded in a host of Polyethylene foam (HD115). The purpose of using a triangular lattice is to ensure isotropic dispersions near the Γ point of the metamaterial [21]. The microstructure of the building block is shown in Fig. 1(a). The cylindrical inclusion is a water cylinder of $r_w = 0.24a$ coated by a layer of silicone rubber of outer radius $r_s = 0.32a$, where *a* is the lattice constant. The presence of soft silicone rubber layer introduces dipolar resonance which gives rise to negative mass density [14]. The water cylinder is chosen to enhance the quadupolar resonance as water can be easily deformed due to its zero shear modulus. The slippery boundary condition has been used on the interface between water and rubber. Comparing to air that was used in previous works [18, 19], the large mass density of water helps to provide the necessary momentum that is needed in any resonance. The material parameters are: $\rho_{water} = 1.0 \times 10^3 \, kg/m^3$, $\lambda_{water} = 2.22 \times 10^9 \, N/m^2$ and $\mu_{water} = 0$ for water, $\rho_{rubber} = 1.3 \times 10^3 \, kg/m^3$, $\lambda_{rubber} = 6.0 \times 10^5 \, N/m^2$, and $\mu_{rubber} = 4.0 \times 10^4 \, N/m^2$ for silicone rubber [12], and $\rho_{foam} = 115 \, kg/m^3$, $\lambda_{foam} = 6.0 \times 10^6 \, N/m^2$, and $\mu_{foam} = 3.0 \times 10^6 \, N/m^2$ for foam [14]. Here λ is the Lamé constant, which is related to the bulk modulus through the relation $\kappa = \lambda + \mu$ in two dimensions. The velocities of longitudinal and transverse waves are defined by $c_l = \sqrt{\kappa + \mu}\sqrt{1/\rho}$ and $c_t = \sqrt{\mu}\sqrt{1/\rho}$, respectively. It can be seen that the wave speeds inside the silicone rubber are much smaller than those in the foam matrix and the water core. These large mismatches at the two interfaces enhance the dipolar and quadrupolar resonances and give rise to a large negative band for shear waves.

The effective medium parameters of the above metamaterial, i.e. effective mass density, $\rho_e$, bulk modulus, $\kappa_e$, and shear modulus, $\mu_e$, can be calculated by using the EMT proposed in [12], which gives: $\frac{\kappa_{foam} - \kappa_e}{\mu_{foam} + \kappa_e} = \frac{4\tilde{D}_0^{ll}}{i\Omega k_{l0}^2}$, $\frac{\rho_{foam} - \rho_e}{\rho_{foam}} = -\frac{8\tilde{D}_1^{ll}}{i\Omega k_{l0}^2}$ and $\frac{\mu_{foam}(\mu_{foam} - \mu_e)}{(\kappa_{foam}\mu_{foam} + (\kappa_{foam} + 2\mu_{foam})\mu_e)} = \frac{4\tilde{D}_2^{ll}}{i\Omega k_{l0}^2}$, where $\Omega$ is the area of the unit cell, $k_{l0}$ is the wave vector of the longitudinal waves in the matrix foam, and $\tilde{D}_m^{ll}$ represent the Mie-like scattering coefficients of the inclusion [12]. The results given by the EMT are plotted in Fig. 1(b), in which the black solid, red dashed and blue dotted curves represent $\rho_e$, $\kappa_e$, and $\mu_e$, respectively. Also plotted in Fig. 1(b) is the sum of the effective shear modulus and bulk modulus, i.e., $\kappa_e + \mu_e$, which is shown by the green dashed-dotted curve. The frequency is shown in dimensionless unit $\tilde{f} = fa/c_{t0}$, where $c_{t0}$ is the transverse wave speed in the matrix foam. Fig. 1(b) clearly shows that $\rho_e$ is negative in a large frequency region from $\tilde{f} = 0.149$ to $\tilde{f} = 0.263$. Within this regime, $\mu_e$ is also negative in a large frequency region from $\tilde{f} = 0.167$ (denoted by "A" in Fig. 1(b)) to $\tilde{f} = 0.216$ (denoted by "B" in Fig. 1(b)). This overlap of negative $\rho_e$ and negative $\mu_e$ regimes indicates that the effective refractive index for transverse waves should also be negative within the frequency regime from A to B. It should be noted that $\kappa_e + \mu_e$ also has a small negative value in a regime from $\tilde{f} = 0.208$ to $\tilde{f} = 0.213$, which implies a negative band for longitudinal waves in this small frequency region.

To verify the prediction of the EMT, we have calculated the band structure by using the MST [20]. The band structure along the ΓM direction is plotted in Fig. 2(a). From which we can see a wide negative band (for shear waves) from $\tilde{f} = 0.161$ to $\tilde{f} = 0.218$ (denoted by red dots) as well as a narrow negative band (for longitudinal waves) from $\tilde{f} = 0.213$ to $\tilde{f} = 0.218$ (denoted by blue dots). This is consistent with the effective medium prediction shown in Fig. 1(b). In Fig. 2(b), we plot the transmission spectra for longitudinal and transverse waves passing through a slab sample of thickness $6a$ and width $50a$ by solid (blue) and dashed (red) curves, respectively. The incident wave is generated by passing a plane wave through a slit of width $20a$, and is incident normally onto the sample surface whose normal is along the ΓM direction. The transmission was obtained by integrating the flux along the incident direction at the output surface and dividing it by the integrated flux in the absence of the slab. Fig. 2(b) shows clearly a passing band (transmission close to unity) for pure shear waves from $\tilde{f} = 0.170$ to $\tilde{f} = 0.213$, while a band gap for longitudinal waves. The oscillations inside the band are due to Fabry-Perot resonances. These results agree well with the band structures shown in Fig. 2(a).

Negative refractive index leads to negative refraction. In order to demonstrate this interesting phenomenon, we use a wedge sample of the metamaterial, which is shown schematically in Fig. 3(a). The wedge sample has 90° and 60° corners, and the two perpendicular surfaces of the wedge are chosen to have their normals along the ΓM (horizontal) and ΓK (vertical) directions of the lattice, respectively. A plane transverse wave passing through a slit of width 20a is incident normally on the vertical surface of the wedge along the ΓM direction. It is clearly seen that the wave penetrates through the wedge and is refracted by the oblique surface. In Fig. 4, we can see if the negative refraction angle $\theta = \phi - 30°$ is positive under a positive incident angle $i$, then negative refraction has occurred. To avoid the reflected waves coming from the bottom surface of the wedge due to internal reflections, we added a little absorption to the last few layers at the bottom of the wedge. In the simulation, we choose $\tilde{f} = 0.2$ as our working frequency and calculate both the near-field and far-field energy flux distributions, i.e., time averaged Poynting vector, $j_k = \pi f \, \text{Im}(\sigma_{ik} u_i^*)$, where $u_i$ and $\sigma_{ik}$ are, respectively, the displacement vector and stress tensor obtained from the MST [22]. The results of the MST calculations are shown in Figs. 3(b) and 3(c) for the near field and far field, respectively. Both results show unambiguously negative refraction. Due to the vector nature of elastic wave, the refracted waves in the far field can be separated into transverse and longitudinal components, which are represented by the dashed (red) and solid (black) curves in Fig. 3(c), respectively. It can be seen that the flux distributions reach their peaks at $\phi = 63°$ (or $\theta = 33°$) for transverse waves and at $\phi = 82°$ (or $\theta = 52°$) for longitudinal waves, respectively. The near field in Fig. 3(b), however, cannot be separated easily due to the interference of the longitudinal and transverse waves.

More interestingly, we note that the far-field flux distribution (as can been seen from the full width at half maximum of the peak) of the longitudinal wave is much larger than that of the transverse wave. This indicates a significant amount of mode conversion from transverse to longitudinal waves has occurred on the output surface of the wedge due to negative refraction. We have integrated the two distributions from $\phi = -60°$ to $120°$ and find that the ratio of total outgoing far-field flux of the transverse waves to that of the longitudinal waves is around 11. Such an efficient mode conversion is a unique property of the double-negative elastic metamaterial. In fact, total mode conversion between shear waves and longitudinal waves can occur on an interface between a double-negative elastic metamaterial and a normal elastic medium under certain conditions. For example, consider a transverse plane wave in a double-negative elastic metamaterial of $\rho_1 < 0$ and $\mu_1 < 0$ is incident on the interface with a normal elastic medium, as shown in Fig. 4. It can be immediately seen that only in the case of negative refraction can the mode conversion from transverse waves to longitudinal waves be complete, i.e. 100%, because the displacements of transverse waves ($\vec{u}_t$) and longitudinal waves ($\vec{u}_l$) can only be matched on the interface when the refraction angle is negative. For a given incident angle $\alpha$, with some simple derivation (shown in the Supplemental Material), we find the total conversion (and no reflection) condition to be $-\mu_1 = \mu_2 \tan 2\alpha \tan \alpha$, $\kappa_2 = \mu_2 / \cos 2\alpha$ and $\tan \alpha = \dfrac{k_l}{k_t} = \dfrac{\sqrt{\mu_1/\rho_1}}{\sqrt{(\kappa_2 + \mu_2)/\rho_2}}$. The negative refraction angle is $\beta = \pi/2 - \alpha$. Such a novel phenomenon may have important applications. Similarly, total mode conversion from longitudinal waves to transverse waves can also be realized under certain conditions (also shown in the Supplemental Material). Total mode conversion is not possible on the interface between two normal solids. However, it can occur in a normal solid with a free surface, but in reflected waves [23].

To conclude, we have proposed a type of elastic metamaterial consisting of a triangular lattice of rubber coated water cylinders in a foam matrix. Both the EMT and the band structure calculation show the existence of a negative band for shear waves in a large frequency region induced by simultaneous negative mass density and shear modulus, which are produced by the enhanced dipolar and quadrupolar resonances, respectively. The absence of longitudinal band in the frequency region makes the metamaterial a good candidate as a polarization filter. Negative refraction has been demonstrated by a wedge sample. Accompanying negative refraction, we also observed a significant amount of mode conversion from transverse waves to longitudinal waves, which is proved to be a unique property of double-negative elastic metamaterials.

The work is supported by Hong Kong RGC Grant No.605008.

**References**

1. V. G. Veselago, *Sov. Phys. Usp.*, **10,** 509 (1968).

2. J. B. Pendry, *Phys. Rev. Lett.* **85**, 3966 (2000).

3. J. B. Pendry, A. J. Holedn, W. J. Stewart, I. Youngs, *Phys. Rev. Lett.* **76**, 4773 (1996); J. B. Pendry, A. J. Holedn, D. J. Robbins, W. J. Stewart, *IEEE Trans. Microwave Theory Tech.* **47**, 2075 (1999).

4. D.R. Smith, W.J. Padilla, D.C. Vier, S.C. Nemat-Nasser, S. Schultz, *Phys. Rev. Lett.* **84**. 4184 (2000); R.A. Shelby, D.R. Smith, S. Schultz, *Science* **292**, 77 (2001).

5. C. M. Soukoulis, S. Linden, M. Wegener, *Science* **315**, 47 (2007).

6. H. J. Lezec, J. A. Dionne, H. A. Atwater, *Science* **316**, 430 (2007).

7. J. Valentine, S. Zhang, T. Zentgraf, E. Ulin-Avila, D. A. Genov, G. Bartal, X. Zhang, *Nature* **455**, 376 (2008).

8. J. Yao, Z. W. Liu, Y. M. Liu, Y. Wang, C. Sun, G. Bartal, A. M. Stacy, X. Zhang, *Science* **321**, 930 (2008).

9. J. Li and C. T. Chan, *Phys. Rev. E* **70**, 055602 (2004).

10. Y. Q. Ding, Z.Y. Liu, C.Y. Qiu, and J. Shi, *Phys. Rev. Lett.* **99**, 093904 (2007).

11. S. Zhang, L. Yin, and N. Fang, *Phys. Rev. Lett.* **102**, 194301 (2009).

12. S.H. Lee, C.M. Park, Y.M. Seo, Z.G. Wang, and C.K. Kim, *Phys. Rev. Lett.* **104**, 054301(2010).

13. G. W. Milton, *New J. Phys.* **9**, 359 (2007).

14. Z. Liu, X. Zhang, Y. Mao, Y.Y. Zhu, Z. Yang, C.T. Chan, and P. Sheng, *Science* **289**, 1734 (2000).

15. Z. Yang, J. Mei, M. Yang, N.H. Chan and P. Sheng, *Phys. Rev. Lett.* **101** , 204301 (2008).

16. S. H. Lee, C. M. Park, Y. M. Seo, Z. G. Wang, and C. K. Kim, *Phys. Lett. A* **373**, 4464 (2009).

17. N. Fang, D. J. Xi, J.Y. Xu, M. Ambrati, W. Sprituravanich, C. Sun, and X. Zhang, *Nat. Mater.* **5**, 452 (2006).

18. Y. Wu, Y. Lai, and Z. Q. Zhang, *Phys. Rev. B* **76**, 205313 (2007).

19. X. Zhou and G. Hu, *Phys. Rev. B* **79**, 195109 (2009).

20. M. Kafesaki and E. N. Economou, *Phys. Rev. B* **60**, 11993 (1999); J. Mei, Z. Liu, J. Shi, and D. Tian, ibid. **67**, 245107 (2003); J. Mei, Z. Liu, W. Wen, and P. Sheng, *Phys. Rev. Lett.* **96**, 024301 (2006); Y. Lai and Z. Q. Zhang, *Appl. Phys. Lett.* **83**, 3900 (2003).

21. Y. Wu and Z. Q. Zhang, Phys. Rev. B **79**, 195111 (2009).

22. Y. Wu, Y. Lai, Y. Wan, and Z. Q. Zhang, Phys. Rev. B **77**, 125125 (2008).

23. For example, D. Royer and E. Dieulesaint, "*Elastic Waves in Solids I*", (Springer, 1999).

**Figures**

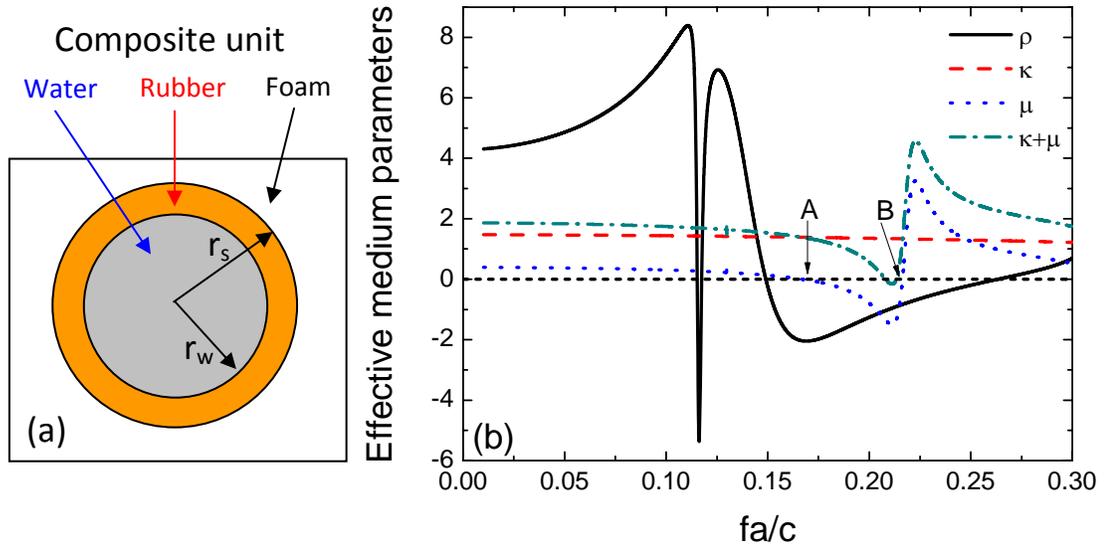

Figure 1 (a) Effective medium parameters obtained from EMT. Solid (black), dashed(red), dotted(blue) and dash-dotted (cyan) correspond to effective mass density, $\rho_e$, bulk modulus, $\kappa_e$, shear modulus, $\mu_e$, and the sum of bulk and shear modulus, $\kappa_e + \mu_e$, respectively. (b) A schematic figure of the building block of the metamaterial: a rubber coated water cylinder embedded in a foam host. The coated cylinder has a radius of $0.32a$ and the radius of the inner water cylinder is $0.24a$, where $a$ is the lattice constant

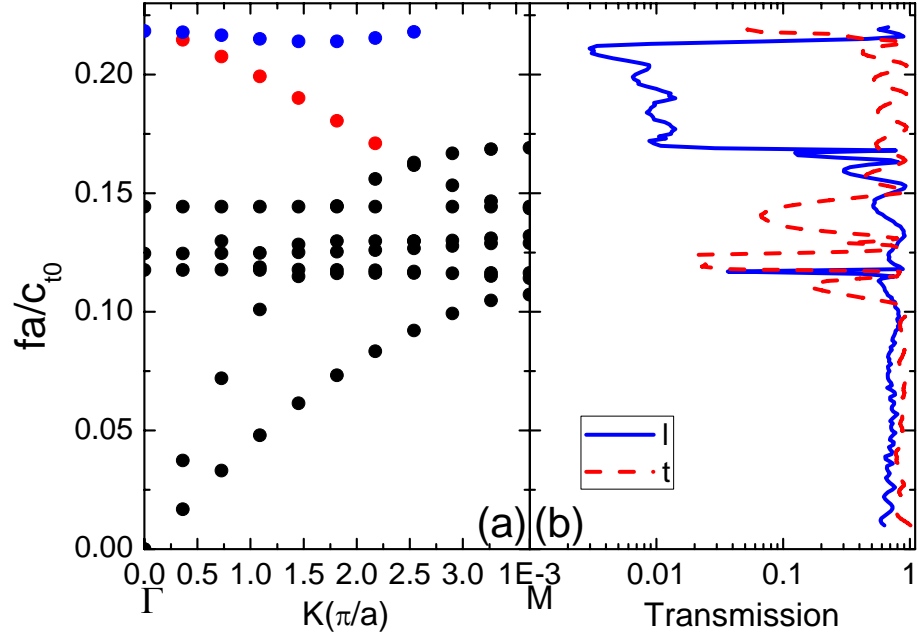

Figure 2 (a) Band structure (along Γ-M direction) of a triangular array of rubber-coated water cylinders embedded in foam host as shown in Fig. 1(b). (b) The solid curve (black) is the transmission coefficient for longitudinal wave incidence and the dashed curve (red) is for transverse wave incidence.

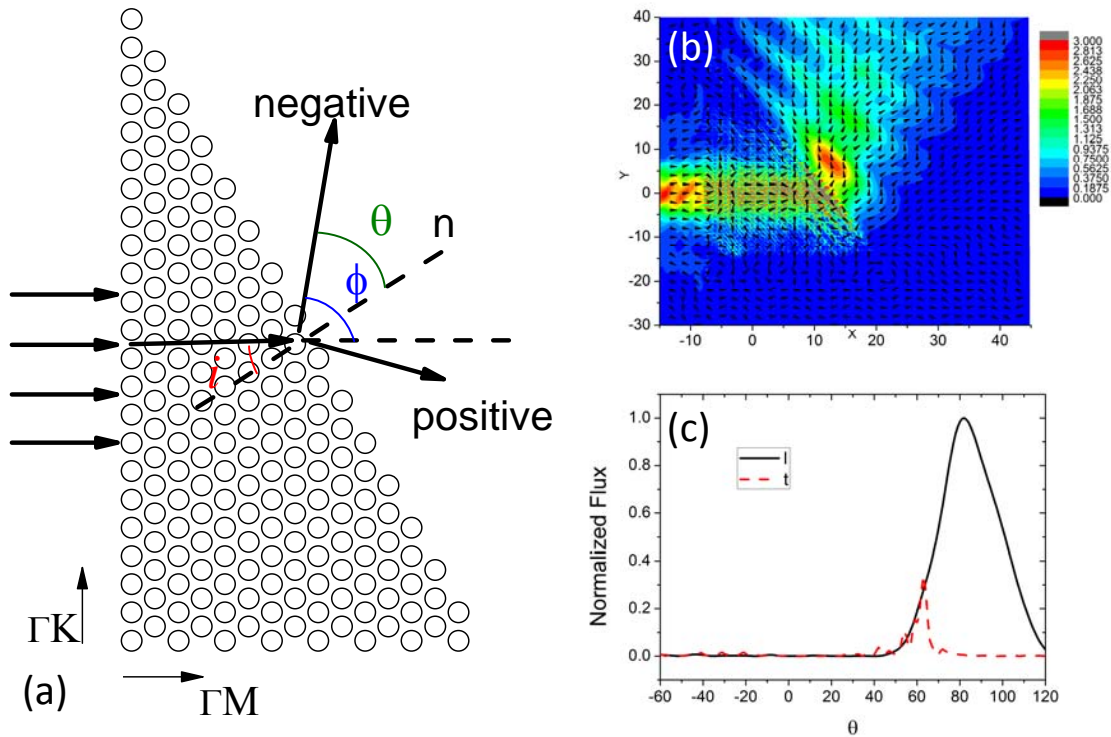

Figure 3  (a) Schematic figure of the sample.  $i$ and $\theta$ are incident and refracted angles respectively. n is the surface normal of the interface between the metamaterial and the background medium. The $\Gamma M$ and $\Gamma K$ directions are along horizontal and vertical directions, respectively.  Incident wave is coming from left shooting normally onto the vertical surface of the wedge. (b) Total flux distribution for a transverse wave incidence on the wedge.  Color and arrows represent the magnitude and directions, respectively.   (c) The far field total flux distribution for the transverse wave incidence. Solid (black) and dashed (red) correspond to longitudinal and transverse waves, respectively.

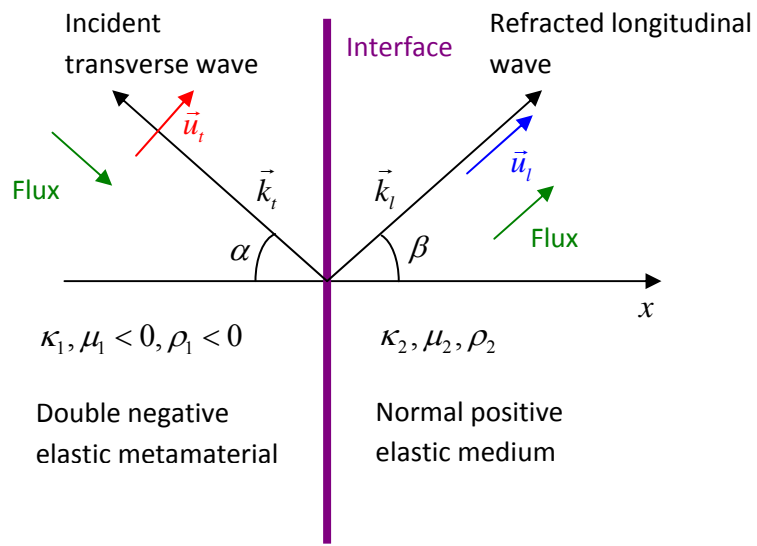

Figure 4. A schematic graph of s-p mode conversion during the negative refraction on the interface between a double-negative elastic metamaterial and a normal elastic medium.

# Supplemental Material for "Elastic metamaterials with simultaneously negative effective shear modulus and mass density"

Ying Wu[1,2], Yun Lai[1,3] and Zhao-Qing Zhang[1*]

[1] Department of Physics, Hong Kong University of Science and Technology, Clear Water Bay, Kowloon, Hong Kong, China

[2] Division of Mathematical and Computer Sciences and Engineering, King Abdullah University of Science and Technology, Thuwal 23955-6900, Kingdom of Saudi Arabia

[3] Department of Physics, Soochow University, 1 Shizi Street, Suzhou 215006, People's Republic of China

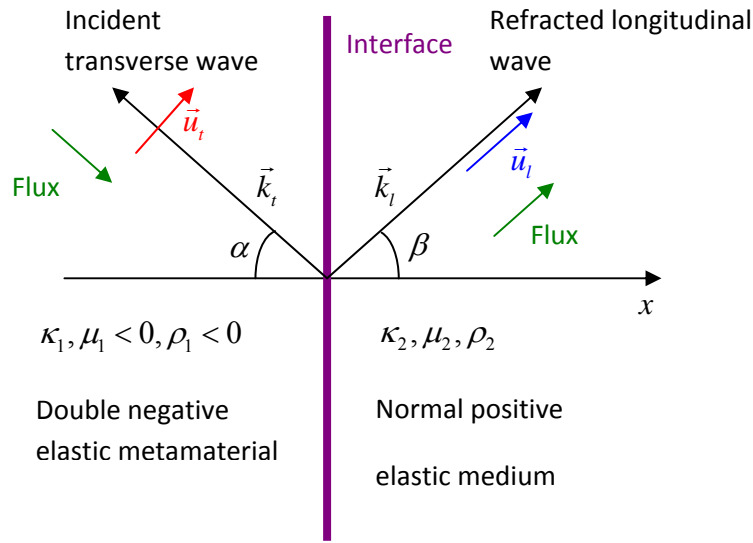

Fig. S1. A schematic graph of *s-p* mode conversion during the negative refraction on the interface of a double-negative elastic metamaterial of Case 1 below.

There can be four cases of total *s-p* or *p-s* mode conversions on the interface of a double-negative metamaterial and a normal solid under negative refraction:

Case 1: Incident shear waves from a double negative metamaterial.

Case 2: Incident longitudinal waves from a double negative metamaterial.

Case 3: Incident shear waves from a normal solid.

Case 4: Incident longitudinal waves from a normal solid.

It is obvious that Case 3 can be obtained from the time reversal of Case 2 and Case 4 can be obtained from the time reversal of Case 1. Below we derive the general conditions of total mode conversions under negative refraction.

For *s* waves, the displacement is $\vec{u}_t = u_t (\sin\alpha \hat{x} + \cos\alpha \hat{y}) e^{ik_t(-\cos\alpha x + \sin\alpha y)}$. For *p* waves, the displacement is $\vec{u}_l = u_l (\cos\beta \hat{x} + \sin\beta \hat{y}) e^{ik_l(\cos\beta x + \sin\beta y)}$. The displacements must match on the interface $x = 0$, i.e.,

$$u_t \sin\alpha e^{ik_t \sin\alpha y} = u_l \cos\beta e^{ik_l \sin\beta y},$$
$$u_t \cos\alpha e^{ik_t \sin\alpha y} = u_l \sin\beta e^{ik_l \sin\beta y}. \tag{S1}$$

From Eq. (S1), we find $u_t \sin\alpha = u_l \cos\beta$ and $u_t \cos\alpha = u_l \sin\beta$, which means that $u_l = u_t$ and $\alpha + \beta = \pi/2$. In the case of normal refraction, for incident angle $0 < \alpha < \pi/2$ we have $\beta < 0$, thus $\alpha + \beta < \pi/2$, the displacements are not possible to match on the interface. Total conversion is only possible in the case of negative refraction.

From Eq. (S1), we also find $k_t \sin\alpha = k_l \sin\beta$, which means the wave vector parallel to the interface must conserve. Substituting $\alpha + \beta = \pi/2$, we obtain $k_t \sin\alpha = k_l \cos\alpha$.

From $k_t = \left|\dfrac{\omega}{v_t}\right| = \dfrac{\omega}{\sqrt{\mu_1/\rho_1}}$ and $k_l = \left|\dfrac{\omega}{v_l}\right| = \dfrac{\omega}{\sqrt{(\kappa_2 + \mu_2)/\rho_2}}$, we find $\tan\alpha = \dfrac{k_l}{k_t} = \dfrac{\sqrt{\mu_1/\rho_1}}{\sqrt{(\kappa_2 + \mu_2)/\rho_2}}$.

We also need to consider the matching condition of stresses. For *s* waves, $\vec{u}_t = u_t (\sin\alpha \hat{x} + \cos\alpha \hat{y}) e^{ik_t(-\cos\alpha x + \sin\alpha y)}$. Thus, the strains are

$$S_{xx}^t = \frac{\partial u_{tx}}{\partial x} = u_t \sin\alpha(-ik_t \cos\alpha)e^{ik_t(-\cos\alpha x + \sin\alpha y)} = -iu_t k_t \sin\alpha \cos\alpha e^{ik_t(-\cos\alpha x + \sin\alpha y)},$$

$$S_{yy}^t = \frac{\partial u_{ty}}{\partial y} = u_t \cos\alpha(ik_t \sin\alpha)e^{ik_t(-\cos\alpha x + \sin\alpha y)} = iu_t k_t \sin\alpha \cos\alpha e^{ik_t(-\cos\alpha x + \sin\alpha y)},$$

$$S_{xy}^t = \frac{1}{2}\left(\frac{\partial u_{tx}}{\partial y} + \frac{\partial u_{ty}}{\partial x}\right)$$

$$= \frac{1}{2}\left(u_t \sin\alpha(ik_t \sin\alpha)e^{ik_t(-\cos\alpha x + \sin\alpha y)} + u_t \cos\alpha(-ik_t \cos\alpha)e^{ik_t(-\cos\alpha x + \sin\alpha y)}\right)$$

$$= \frac{1}{2}\left(iu_t k_t \sin^2\alpha - iu_t k_t \cos^2\alpha\right)e^{ik_t(-\cos\alpha x + \sin\alpha y)}.$$

And the stress can be written as

$$\sigma_{xx}^t = (\kappa_1 + \mu_1)S_{xx}^t + (\kappa_1 - \mu_1)S_{yy}^t$$
$$= -(\kappa_1 + \mu_1)iu_t k_t \sin\alpha \cos\alpha e^{ik_t(-\cos\alpha x + \sin\alpha y)} + (\kappa_1 - \mu_1)iu_t k_t \sin\alpha \cos\alpha e^{ik_t(-\cos\alpha x + \sin\alpha y)}$$
$$= -2i\mu_1 u_t k_t \sin\alpha \cos\alpha e^{ik_t(-\cos\alpha x + \sin\alpha y)}$$
$$= -i\mu_1 u_t k_t \sin 2\alpha e^{ik_t(-\cos\alpha x + \sin\alpha y)},$$

$$\sigma_{yy}^t = (\kappa_1 + \mu_1)S_{yy}^t + (\kappa_1 - \mu_1)S_{xx}^t$$
$$= (\kappa_1 + \mu_1)iu_t k_t \sin\alpha \cos\alpha e^{ik_t(-\cos\alpha x + \sin\alpha y)} - (\kappa_1 - \mu_1)iu_t k_t \sin\alpha \cos\alpha e^{ik_t(-\cos\alpha x + \sin\alpha y)}$$
$$= 2i\mu_1 u_t k_t \sin\alpha \cos\alpha e^{ik_t(-\cos\alpha x + \sin\alpha y)}$$
$$= i\mu_1 u_t k_t \sin 2\alpha e^{ik_t(-\cos\alpha x + \sin\alpha y)},$$

$$\sigma_{xy}^t = 2\mu_1 S_{xy}^t$$
$$= \mu_1\left(iu_t k_t \sin^2\alpha - iu_t k_t \cos^2\alpha\right)e^{ik_t(-\cos\alpha x + \sin\alpha y)}$$
$$= -i\mu_1 u_t k_t \cos 2\alpha e^{ik_t(-\cos\alpha x + \sin\alpha y)}.$$

For $p$ waves, $\vec{u}_l = u_l(\cos\beta \hat{x} + \sin\beta \hat{y})e^{ik_l(\cos\beta x + \sin\beta y)}$. Thus, the strains are

$$S_{xx}^l = \frac{\partial u_{lx}}{\partial x} = u_l \cos\beta(ik_l \cos\beta)e^{ik_l(\cos\beta x + \sin\beta y)} = iu_l k_l \cos^2\beta e^{ik_l(\cos\beta x + \sin\beta y)},$$

$$S_{yy}^l = \frac{\partial u_{ly}}{\partial y} = u_l \sin\beta(ik_l \sin\beta)e^{ik_l(\cos\beta x + \sin\beta y)} = iu_l k_l \sin^2\beta e^{ik_l(\cos\beta x + \sin\beta y)},$$

$$S_{xy}^l = \frac{1}{2}\left(\frac{\partial u_{lx}}{\partial y} + \frac{\partial u_{ly}}{\partial x}\right)$$

$$= \frac{1}{2}\left(u_l \cos\beta(ik_l \sin\beta)e^{ik_l(\cos\beta x+\sin\beta y)} + u_l \sin\beta(ik_l \cos\beta)e^{ik_l(\cos\beta x+\sin\beta y)}\right)$$

$$= iu_l k_l \sin\beta \cos\beta e^{ik_l(\cos\beta x+\sin\beta y)}.$$

And the stress can be written as

$$\sigma_{xx}^l = (\kappa_2 + \mu_2)S_{xx}^l + (\kappa_2 - \mu_2)S_{yy}^l$$
$$= (\kappa_2 + \mu_2)iu_l k_l \cos^2\beta e^{ik_l(\cos\beta x+\sin\beta y)} + (\kappa_2 - \mu_2)iu_l k_l \sin^2\beta e^{ik_l(\cos\beta x+\sin\beta y)}$$
$$= i\left(\kappa_2(\cos^2\beta + \sin^2\beta) + \mu_2(\cos^2\beta - \sin^2\beta)\right)u_l k_l e^{ik_l(\cos\beta x+\sin\beta y)}$$
$$= i(\kappa_2 + \mu_2 \cos 2\beta)u_l k_l e^{ik_l(\cos\beta x+\sin\beta y)},$$

$$\sigma_{yy}^l = (\kappa_2 + \mu_2)S_{yy}^l + (\kappa_2 - \mu_2)S_{xx}^l$$
$$= (\kappa_2 + \mu_2)iu_l k_l \sin^2\beta e^{ik_l(\cos\beta x+\sin\beta y)} + (\kappa_2 - \mu_2)iu_l k_l \cos^2\beta e^{ik_l(\cos\beta x+\sin\beta y)}$$
$$= i\left(\kappa_2(\cos^2\beta + \sin^2\beta) + \mu_2(-\cos^2\beta + \sin^2\beta)\right)u_l k_l e^{ik_l(\cos\beta x+\sin\beta y)}$$
$$= i(\kappa_2 - \mu_2 \cos 2\beta)u_l k_l e^{ik_l(\cos\beta x+\sin\beta y)},$$

$$\sigma_{xy}^l = 2\mu_2 S_{xy}^l$$
$$= 2\mu_2 iu_l k_l \sin\beta \cos\beta e^{ik_l(\cos\beta x+\sin\beta y)}$$
$$= i\mu_2 u_l k_l \sin 2\beta e^{ik_l(\cos\beta x+\sin\beta y)}.$$

$\sigma_{xx}$ and $\sigma_{xy}$ must also be continuous on the interface $x = 0$, i.e.,

$$-i\mu_1 u_t k_t \sin 2\alpha e^{ik_t \sin\alpha y} = i(\kappa_2 + \mu_2 \cos 2\beta)u_l k_l e^{ik_l \sin\beta y},$$
$$-i\mu_1 u_t k_t \cos 2\alpha e^{ik_t \sin\alpha y} = i\mu_2 u_l k_l \sin 2\beta e^{ik_l \sin\beta y}.$$
(S2)

By substituting $\alpha + \beta = \pi/2$, $u_l = u_t$ and $k_t \sin\alpha = k_l \cos\alpha$ into Eq. (S2), we find

$$\kappa_2 = \mu_2 \cos 2\alpha - \mu_1 \sin 2\alpha/\tan\alpha \text{ and } -\mu_1 = \mu_2 \tan 2\alpha \tan\alpha.$$

Conclusion: total conversion is possible under the case of negative refraction.

The total conversion matching conditions are:

$$\alpha + \beta = \pi/2,$$
$$\kappa_2 = \mu_2 \cos 2\alpha - \mu_1 \sin 2\alpha / \tan \alpha,$$
$$-\mu_1 = \mu_2 \tan 2\alpha \tan \alpha, \qquad (S3)$$
$$\tan \alpha = \frac{k_l}{k_t} = \frac{\sqrt{\mu_1/\rho_1}}{\sqrt{(\kappa_2 + \mu_2)/\rho_2}}.$$

Since Case 3 can be obtained from the time reversal of Case 2 and Case 4 can be obtained from the time reversal of Case 1, here we discuss only Cases 1 and 3.

Case 1: we consider a transverse plane wave incident from the left medium of $\mu_1 < 0, \rho_1 < 0$ with an incident angle of $0 < \alpha < \pi/2$ as shown in Fig. S1. First, we can obtain $\mu_2$ by using $-\mu_1 = \mu_2 \tan 2\alpha \tan \alpha$. Then, we can obtain $\kappa_2$ by using $\kappa_2 = \mu_2 \cos 2\alpha - \mu_1 \sin 2\alpha / \tan \alpha = \mu_2/\cos 2\alpha$. At last, we can obtain $\rho_2$ by using $\tan \alpha = \frac{k_l}{k_t} = \frac{\sqrt{\mu_1/\rho_1}}{\sqrt{(\kappa_2 + \mu_2)/\rho_2}}$. Therefore, $\mu_2$, $\kappa_2$ and $\rho_2$ are all obtained. It is worth mentioning that when $\alpha > \pi/4$, we have $\mu_2 < 0$, but $\kappa_2 > 0$ and $\kappa_2 + \mu_2 = \mu_2(1 + \cos 2\alpha)/\cos 2\alpha > 0$, which indicates a double positive medium for refracted longitudinal waves on the right (together with $\rho_2 > 0$). In this case the medium on the right is not a normal solid.

Case 3: we consider a transverse plane wave incident from the left medium of $\mu_1 > 0, \rho_1 > 0$ with an incident angle of $0 < \alpha < \pi/2$. $\mu_2$, $\kappa_2$ and $\rho_2$ can also be obtained from Eq. (S3). Note that we have $\kappa_2 + \mu_2 = \mu_2(1 + \cos 2\alpha)/\cos 2\alpha < 0$ and $\rho_2 < 0$, which indicate a double negative medium for refracted longitudinal waves on the right.